\documentstyle[12pt,epsf]{article}

\newlength{\pubnumber} 

\catcode`\@=11
\@addtoreset{equation}{section}

\def\section{\@startsection{section}{1}{\z@}{3.5ex plus 1ex minus .2ex}
 {2.3ex plus .2ex}{\large\bf}}
\def\subsection{\@startsection{subsection}{2}{\z@}{2.3ex plus .2ex}
 {2.3ex plus .2ex}{\bf}}

\def\beq{\begin{equation}}
\def\eeq{\end{equation}}
\def\beqn{\begin{eqnarray}}
\def\eeqn{\end{eqnarray}}

\def\nolabel{\nonumber }

\def\half{{\textstyle{1\over 2}}}
\def\third{{\textstyle {1\over3}}}
\def\fourth{{\textstyle {1\over4}}}

\def\sixth{{\textstyle {1\over6}}}
\def\tenth{{\textstyle {1\over10}}}

\def\twothird{{\textstyle {2\over3}}}
\def\fivethird{{\textstyle {5\over3}}}
\def\threefourth{{\textstyle {3\over4}}}

\def\MS{M_{str}}

\def\MP{M_{PL}}

\def\bone{{\mathbf 1}}

\def\mb{{\mathbf b}}

\def\p4{\Phi_4}
\def\pp4{\Phi^{'}_4}
\def\pb4{\bar{\Phi}_4}
\def\ppb4{\bar{\Phi}^{'}_4}
\def\p#1{{\Phi_{#1}}}

\def\pp#1{{\Phi^{'}_{#1}}}
\def\pb#1{{{\overline{\Phi}}_{#1}}}

\def\ppb#1{{{\overline{\Phi}}^{'}_{#1}}}

\def\FD2pv{FD2$^{'}$V }
\def\FD2p{FD2$^{'}$ }

\def\inbar{\,\vrule height1.5ex width.4pt depth0pt}

\def\IC{\relax\hbox{$\inbar\kern-.3em{\rm C}$}}
\def\IQ{\relax\hbox{$\inbar\kern-.3em{\rm Q}$}}
\def\IR{\relax{\rm I\kern-.18em R}}
 \font\cmss=cmss10 \font\cmsss=cmss10 at 7pt

\def\IZ{\relax\ifmmode\mathchoice
 {\hbox{\cmss Z\kern-.4em Z}}{\hbox{\cmss Z\kern-.4em Z}}
 {\lower.9pt\hbox{\cmsss Z\kern-.4em Z}}
 {\lower1.2pt\hbox{\cmsss Z\kern-.4em Z}}\else{\cmss Z\kern-.4em Z}\fi}

\def\Io{\relax\ifmmode\mathchoice
 {\hbox{\cmss 1\kern-.4em 1}}{\hbox{\cmss 1\kern-.4em 1}}
 {\lower.9pt\hbox{\cmsss 1\kern-.4em 1}}
 {\lower1.2pt\hbox{\cmsss 1\kern-.4em 1}}\else{\cmss 1\kern-.4em 1}\fi}

\def\u{\underline{\phantom{a}}}
\begin{document}

\begin{titlepage}
\samepage{
\setcounter{page}{1}
\rightline{BU-HEPP-02/02}
\rightline{CASPER-02/02}
\rightline{\tt hep-ph/0209050}
\rightline{August 2002}
\vfill
\begin{center}
 {\Large \bf  On the Possibility of Optical Unification\\
                    in Heterotic Strings\\
}
\vfill
\vskip .4truecm
\vfill {\large
        G. Cleaver,$^{1,2}$\footnote{Gerald$\u$Cleaver@baylor.edu}
        V. Desai,$^{1,3}$\footnote{verlox7516@yahoo.com} 
        H. Hanson,$^{1,4}$\footnote{hah1@students.uwf.edu} 
        J. Perkins,$^{1}$\footnote{John$\u$Perkins@baylor.edu}\\ 
        D. Robbins,$^{1,5}$\footnote{drobbin@tulane.edu}  
     \& S. Shields$^{1,6}$\footnote{sshields@calpoly.edu}
}
\\
\vspace{.12in}
{\it $^{1}$ Center for Astrophysics, Space Physics \& Engineering Research,\\
            Dept.\ of Physics, PO Box 97316, Baylor University,\\
            Waco, TX 76798-7316, USA\\}
\vspace{.06in}
{\it $^{2}$ Astro Particle Physics Group,
            Houston Advanced Research Center (HARC),\\
            The Mitchell Campus,
            Woodlands, TX 77381, USA\\}
\vspace{.06in}
{\it $^{3}$ St. Thomas Episcopal School,\\
            4900 Jackwood, Houston, TX 77096, USA,\\}
\vspace{.06in}
{\it $^{4}$ Department of Physics,
            University of West Florida,\\
            Pensacola, FL 32514, USA\\}
\vspace{.06in}
{\it $^{5}$ Department of Physics,
            Tulane University,\\
            New Orleans, LA 70118, USA\\}
\vspace{.06in}
{\it $^{6}$ Department of Physics,
            California Polytechnic State University,\\
            San Luis Obispo, CA 93407, USA\\}
\vspace{.025in}
\end{center}
\vfill
\newpage
\begin{abstract}
Recently J.\ Giedt discussed a mechanism, entitled 
optical unification,  
whereby string scale unification is facilitated via
exotic matter with intermediate scale mass.  
This mechanism guarantees that
a virtual MSSM unification below the string scale 
is extrapolated from the running of gauge couplings upward
from $M_{Z^o}$ 
when an intermediate scale desert is assumed.
In this letter we explore the possibility of optical unification 
within the context of weakly coupled heterotic strings.
In particular, we investigate this for models of free fermionic 
construction containing the NAHE set of basis vectors.
This class is of particular interest for optical unification, because it 
provides a standard hypercharge embedding within $SO(10)$,  
giving the standard $k_Y= \fivethird$ hypercharge level, which was shown
necessary for optical unification.
We present a NAHE model for which 
the set of exotic $SU(3)_C$ triplet/anti-triplet pairs, $SU(2)_L$ doublets, 
and non-Abelian singlets with hypercharge 
offers the possibility of optical unification. 
Whether this model can realize optical unification 
is conditional upon these exotics not receiving
Fayet-Iliopoulos (FI) scale masses 
when a flat direction of scalar vacuum expectation values 
is non-perturbatively chosen to cancel 
the FI $D$-term $\xi$ generated by the anomalous
$U(1)$-breaking
Green-Schwarz-Dine-Seiberg-Wittten mechanism. 
A study of perturbative flat directions and their phenomenological 
implications for this model is underway.

{\it This paper is a product of the NFS 
Research Experiences for Undergraduates  
and the NSF High School Summer Science Research programs 
at Baylor University.} 
\end{abstract}
\smallskip}
\end{titlepage}

\setcounter{footnote}{0}

%==============================================================================
%$SU(3)_C \times SU(2)_L \times U(1)_Y \times \prod_i U(1)_i$.
%============================== SECTION 1 ============================
% at here

\section{Optical Unification}

An enduring issue in string theory has been the discrepancy between the 
$SU(3)_C\times SU(2)_L\times U(1)_Y$ ([321])
gauge coupling unification scale $\Lambda_U$ for 
the Minimal Supersymmetric Standard Model (MSSM) 
with intermediate scale desert 
and the string scale $\Lambda_H$ for  
the weakly coupled heterotic string. 
When couplings are run upward from their values near 
$M_{Z^o}$, a MSSM unification scale 
$\Lambda_U \sim 2.5 \times 10^{16}$ GeV \cite{mssmgcu}
is predicted for an intermediate scale desert.
In contrast, the weakly coupled heterotic string 
scale is around $\Lambda_H \sim 5\times 10^{17}$ \cite{kapl}. 

Three types of solutions have been proposed to 
resolve this factor of 20 disagreement \cite{kd}.
One proposal is a grand unified 
theory between $\Lambda_U$ and $\Lambda_H$. Here the MSSM couplings
merge at $\Lambda_U$ and then run together within a GUT to
$\Lambda_H$.
However, with the exception of flipped $SU(5)$ \cite{fsu5} 
(or partial GUTs such as the Pati-Salam $SU(4)_C\times SU(2)_L\times SU(2)_R$ 
\cite{lrs})  
string GUT models based on level-one Ka\v c-Moody algebras encounter a 
difficulty: they lack the required adjoint higgs (and 
higher representations). 
Alternately, strong coupling effects of 
$M$-theory can lower $\Lambda_H$ down to $\Lambda_U$ \cite{witten}.
Conversely, intermediate scale exotics could shift
the MSSM unification scale upward to the string scale \cite{mup}.

The near ubiquitous appearance of MSSM-charged exotics in 
heterotic string models adds weight to the third 
proposal.\footnote{Even the underlying model from which  
the Minimal Supersymmetric Heterotic String Model 
[MSHSM] is derived contains MSSM-charged exotics \cite{cfn1}. 
These exotics are, however, eliminated from the MSHSM low energy 
effective field theory by a set of anomaly-cancelling flat 
directions \cite{cfn2,cfn3}.} If MSSM exotics exist with 
intermediate scale masses of order $\Lambda_I$, then the actual 
[321] running couplings are altered above $\Lambda_I$. It is then, 
perhaps, phenomenologically puzzling that the illusion of MSSM unification
should still be maintained when the
intermediate scale MSSM exotics are ignored \cite{giedt02a}.  
Maintaining this illusion
likely requires very fine tuning of $\Lambda_I$ for a generic 
exotic particle set and $\Lambda_H$.  
Slight shifting of $\Lambda_I$ would, with high probability, destroy appearances.
Thus, in some sense, the apparent MSSM unification below the 
string scale might be viewed as accidental \cite{ghil,giedt02a}.
 
A mechanism whereby the appearance of 
a $\Lambda_U$ is not accidental would be very appealing. 
Just such a mechanism, entitled ``optical unification,''  
has recently been discussed by Joel Giedt \cite{giedt02a}. 
Optical unification results in $\Lambda_U$ not disappearing 
under shifts of $\Lambda_I$. 
Instead, $\Lambda_U$ likewise shifts in value.
This effect is parallel to a virtual image always appearing 
between a diverging lens and a real object, 
independent of the position of the lens or real object. 
Hence, Giedt's choice of appellation for this mechanism.

Successful optical unification requires three things \cite{giedt02a}. 
First, the effective level of the 
hypercharge generator must be the standard 
\beqn
k_Y = \textstyle{5\over3}\, . 
\label{db123k}
\eeqn
(\ref{db123k}) is a 
strong constraint on string-derived $[321]$ models,
for the vast majority have non-standard 
hypercharge levels \cite{giedt02b}.
Only select classes of models, such as the NAHE-based \cite{nahe}
free fermionic class \cite{fff},
can yield $k_{Y} = \textstyle{5\over3}$.\footnote{The 
``NAHE'' set is named after 
Nanopoulos, Antoniadis, Hagelin and Ellis and refers to a specific set of 
of five free fermionic basis vectors, which first appeared in the 
construction of string-based flipped $SU(5)$.}
Second, optical unification imposes the relationship 
\beqn
\delta b_2 = \textstyle{7\over12} \delta b_3 + \textstyle{1\over4} \delta b_Y \, .
\label{db123}
\eeqn 
between the exotic particle contributions $\delta b_3$, $\delta b_2$, and $\delta b_1$
to the [321] beta function coefficients. 
Each $SU(3)_C$ exotic triplet or anti-triplet contributes 
$\half$ to $\delta b_3$;
each $SU(2)_C$ doublet contributes  
$\half$ to $\delta b_2$.
With the hypercharge of a MSSM quark doublet normalized to $\sixth$,
the contribution to $\delta b_Y$ from an individual particle with 
hypercharge $Q_Y$ is $Q_Y^2$. 
$\delta b_3 > \delta b_2$ is required 
to keep the virtual unification scale below the string scale. 
Combining this with (\ref{db123}) imposes
\beqn
\delta b_3 >  \delta b_2 \ge \textstyle{7\over12} \delta b_3 \, ,
\label{db123b}
\eeqn 
since $\delta b_Y \geq 0$. 

To acquire intermediate scale mass, 
the exotic triplets and anti-triplets must be equal in number.  
Similarly, the exotic doublets must be even in number. 
Hence, $\delta b_3$ and $\delta b_2$ must be integer \cite{giedt02a}. 
As Giedt pointed out, the simplest solution to
(\ref{db123}) and (\ref{db123b}) is a set of 
three exotic triplet/anti-triplet pairs and two pairs of doublets. 
One pair of doublets can carry $Q_Y=\pm \half$, while the remaining
exotics carry no hypercharge. 
Alternately, if the doublets carry too little hypercharge,
some exotic $SU(3)_C \times SU(2)_L$ singlets could make up the 
hypercharge deficit. 
The next simplest 
solution requires four triplet/anti-triplet pairs and three pairs of 
doublets that yield $\delta b_Y = 2 \textstyle{2\over 3}$
either as a set, or with the assistance of additional non-Abelian singlets. 
For more than four $3/\bar{3}$ pairs, (\ref{db123})
and (\ref{db123b}) allow more than one number of doublet pairs.

\section{Possible Optical Unification in Heterotic Strings}

An excellent class of string models to investigate 
for possible optical unification is
heterotic models of free fermionic construction containing the NAHE set.  
In particular, 
these models can always provide a standard $k_Y= \fivethird$ 
hypercharge embedding \cite{fsu5,fny,af2,af3}. 
(\ref{db123k}) is a very non-trivial constraint which
generally cannot be satisfied for models of other classes \cite{giedt02b}. 
Thus, we re-examined the known NAHE-based $[321]$ models 
\cite{fny,af2,af3,af3a,af3b,af3c} 
and $[322]$ left-right symmetric extensions \cite{cfs,ccf}   
with regard to their MSSM exotic particle content.

Generic NAHE-based [321] models contain a only few exotic $SU(3)_C$ 
triplet/anti-triplet pairs.  
The models of \cite{fny}, \cite{af2}, and \cite{af3}  
contain one pair, two pairs, and one pair, respectively 
The models of \cite{af3a}, \cite{af3b}, and Table 5 model of \cite{af3c}  
each contain 3 pairs.
The [322] left-right symmetric models 1, 2, and 3 of \cite{cfs} 
contain one, four, and two pairs, while the model 
of \cite{ccf} contains three pairs.   
Thus, the models of \cite{fny}, \cite{af2}, and \cite{af3}, along with
models 1 and 3 of \cite{cfs}, contain too few
$SU(3)$ exotics to generate optical unification.    

The basis vectors of the \cite{af3}, \cite{af3a}, and \cite{af3b}
models are the same. The variations among the corresponding models are 
solely a function of some differing GSO phase choices. 
Noting this, we have investigated further variations in GSO phase choices 
for these basis vectors.\footnote{Because of the symmetries 
within the basis vectors generating these models, 
differing GSO phase choices can yield identical models. The complete set of 
models produced from the distinct GSO phase choices 
for the basis vectors of \cite{af3} will be discussed elsewhere.}
Our investigation has resulted in five additional models, denoted
FCREU1 through FCREU5, which respectively 
contain four pairs, three pairs, two pairs, no pairs, 
and, again, no pairs of exotic triplets/anti-triplets. 
The gauge groups and MSSM-charged exotic particle content of these models
appear in Table A1. 
The corresponding GSO phase variations 
from those of \cite{af3} are listed in Table A2.  

Of the [321] models with three or more exotic $3/\bar{3}$ pairs, 
we focus on model FCREU1 with its 4 exotic pairs. 
(In Table A3 we list the complete set of states of this model and their charges.)
Each of the other models presents difficulties:
The model of \cite{af3b} contains too few exotic
doublets (only three rather than four) to satisfy (\ref{db123b}).  
The model FCREU2 with three exotic triplet/anti-triplet pairs 
contains nine exotic $SU(2)_L$ doublets, but six of these doublets
are coupled in pairs via a custodial $SU(2)_C$. 
Further, with regard to MSSM exotics, model FCREU1 
is an enhanced version of the model in \cite{af3a}.  
Any optical solution involving exactly three triplet/anti-triplet pairs from model FCREU1    
is also a solution for the model of \cite{af3a} (and vice versa).  
The left-right symmetric models of \cite{cfs} and \cite{ccf} with three or 
more exotic $3/\bar{3}$ pairs will be discussed in a separate paper.

The MSSM-charged exotics of model FCREU1 are    
four $SU(3)_C$ $3/\bar{3}$ pairs,
three $SU(2)_L$ doublet pairs, 
and seven pairs of MSSM exotic non-Abelian singlets carrying $Q_Y= \pm\half$, 
in addition to three pairs of linear combinations of 
the MSSM higgs-like doublets, $h_{i=1\, {\rm to} \, 4}$ and  
$\bar{h}_{i=1\, {\rm to} \, 4}$.
The extra higgs-like doublets are generally present in NAHE-based [321] 
models. 
Three of the $SU(3)_C$ exotic pairs carry $Q_Y= \pm\third$, while the remaining pair 
carries $Q_Y= \pm\sixth$. None of the exotic $SU(2)_L$ non-higgs doublets 
carry hypercharge. 
For successful optical unification,
when all four $3/\bar{3}$ exotic pairs receive intermediate scale masses, 
all three pairs of exotic non-higgs doublets must also. 
Further, the total contribution to $\delta b_Y$ from all 
$3/\bar{3}$ pairs is $2 \sixth$, while (\ref{db123}) requires a total
$\delta b_Y= 2 \twothird$. Since the  
exotic non-higgs doublets have no hypercharge, the
remainder of $2 \twothird - 2 \sixth = \half$ must be provided by  
exactly two exotic singlets, 
each contributing $\fourth$ to $\delta b_Y$. 
The six remaining pairs of exotic singlets must 
acquire Fayet-Iliopoulos (FI) scale masses.

If one exotic $3/\bar{3}$ pair receives near string scale mass, leaving 
only three exotic $3/\bar{3}$ pairs to receive intermediate scale masses,
the total triplet/anti-triplet contribution to 
$\delta b_Y$ can be 2, $1 \threefourth$, or $1 \half$, 
depending on whether zero, one, or two intermediate scale
(anti-)triplets carry $Q_Y= \pm\sixth$, respectively. 
Even the lowest of these choices provides too 
large of a contribution to $\delta b_Y$, 
since for three exotic $3/\bar{3}$ pairs 
(\ref{db123}) and (\ref{db123b}) require $\delta b_Y= 1$. 
Hence, optical unification cannot be achieved for this model 
when only three $3/\bar{3}$ pairs acquire intermediate scale 
masses. This also implies optical unification is not possible for the model of 
\cite{af3a}.
The Table 5 model of \cite{af3c} is another with
three exotic $3/\bar{3}$ pairs, of which each has $Q_Y=\pm\third$.
Thus, the Table 5 model is similarly prohibited from optical unification.        
 
Optical lensing might be possible in other models 
containing exactly 
three hypercharged exotic $3/\bar{3}$ pairs, if these pairs were of the 
$Q_Y= \pm\sixth$ class.
Then their total contribution to $\delta b_Y$ would be $\half$. 
The remaining contribution of 
$\half$ for the required total $\delta b_Y= 1$ 
could be provided by two singlets, each with $Q_Y= \pm\half$. 
A $3$ or $\bar{3}$ with $Q_Y= \pm\third$ originates in a sector for which
the fermions generating the $SU(3)_C$ symmetry have
antiperiodic boundary conditions, while 
a $3$ or $\bar{3}$ with $Q_Y= \pm\sixth$ originates in a sector for which
these fermions have periodic boundary conditions.  
Whether or not a model can have exactly three exotic $3/\bar{3}$ pairs 
with $Q_Y= \pm\sixth$ is under investigation \cite{cfmnpw}. 
 
In addition to containing the set of four $3/\bar{3}$ pairs, 
three non-higgs-like doublet pairs, and one pair of singlets, which
satisfies the optical unification 
constraints (\ref{db123k})--(\ref{db123b}), model FCREU1 possesses 
the other six pairs of $Q_Y = \pm\half$ singlets and 
the three extra higgs-like doublets. 
For successful optical unification these additional MSSM exotics
must receive near string scale masses, while the optical unifying
set does not. 

Like the other NAHE-based standard-like models,
the model of \cite{af3}, and its variations in \cite{af3a,af3b,af3c}
and herein, contain an anomalous $U(1)$ 
\cite{anomu1}.\footnote{A few semi-realistic non-anomalous 
NAHE-based models have been constructed, 
such as the first two left-right symmetric models in \cite{cfs},
but these have enhanced observable sector symmetries.}
The anomalous Abelian symmetry is broken by the
Green-Schwarz-Dine-Seiberg-Witten mechanism \cite{dsw}, 
which generates a contribution, $\xi$, to the anomalous FI $D$-term in the process.  
A flat direction of vacuum expectation values (VEVs) of scalars  
is then non-perturbatively chosen to  
cancel the FI $\xi$--term, 
restoring supersymmetry and stabilizing the vacuum.
The FI VEV scale is typically $\sim \tenth$ of the string scale.
Superpotential interaction of the flat direction VEVs with superfields
can generate (near) FI scale masses for various states. 
Thus, ideally it may be possible for 
the six additional pairs of exotic singlets 
and three pairs of extra higgs-like doublets to receive FI scale masses
from the flat direction VEVs, while the optical unification
exotics remain massless at the FI scale.

Singlet and non-singlet flat directions of the MSHSM 
were constructed in \cite{cfn1,cfn2,cfn3}
that give FI scale masses to exactly three out of four pairs 
of higgs-like doublets. 
For these flat directions, the physical higgs doublets 
$h$ and $\bar{h}$ are formed from a linear 
combination of three or more $h_{i}$ and $\bar{h_i}$, respectively, 
while three orthogonal linear combinations of $h_{i}$ and of $\bar{h_i}$ 
receive FI scale mass. Similar flat directions
may accomplish this for model FCREU1 also. 
Such flat directions must also generate FI scale masses 
for exactly six of the seven pairs of exotic $Q_Y$-carrying singlets. 
Interestingly, as Table A3 indicates, the pair of singlets denoted
$A_1/\bar{A}_1$ does not follow the charge pattern of the other six. 
This is the singlet pair most likely to remain massless at the FI 
scale.
To be consistent with optical unification, a flat direction must, of course, 
keep the four pairs of exotic MSSM triplets and the three pairs of 
exotic non-higgs-like doublets FI scale massless, as we have discussed. 
Studies of model FCREU1 perturbative flat directions and of
their implications regarding masses are underway \cite{cfmnpw}. 
The methods of this investigation are parallel to those followed in
the flat direction studies of the MSHSM \cite{cfn1,cfn2,cfn3}, 
flipped SU(5) \cite{cen1,cenmw1},
and of other standard-like and semi-GUT models \cite{cfs,ccf,cfv,cnpw} 
located in the parameter space of NAHE-based models.  
 
Assuming that the required exotic triplets, doublets, and non-Abelian
singlets remain massless at the FI scale, an intermediate mass scale 
must be generated for them, perhaps through 
ninth or higher order mass terms resulting from flat direction VEVs or
hidden sector condensation. Mass terms for specific flat 
directions will be studied in \cite{cfmnpw}. 
Alternately, for a generic $SU(N_c)$ gauge group containing $N_f$ flavors
of matter states in vector-like pairings
$H_i {\bar H}_i$, $i= 1,\, \dots\, N_f$,
the gauge coupling $g_i$,
though weak at the string scale $\MS$, becomes strong
for $N_f < 3 N_c$ at a condensation scale defined by
\beqn
\Lambda = \MP {\rm e}^{8 \pi^2/\beta g_s^2}\, ,
\label{consca}
\eeqn
where the $\beta$--function is given by,
\beqn
\beta = - 3 N_c + N_f\, .
\label{befn}
\eeqn
The $N_f$ flavors counted are only those that ultimately receive
masses $m\ll \Lambda$. The hidden sector matter states of model FCREU1 
are 
four $5/\bar{5}$ pairs of $SU(5)_H$ and   
four $3/\bar{3}$ pairs of $SU(3)_H$. Any number of these states 
from none to all might become FI scale massive under a flat direction. 
Thus, $\beta_5$ could be anywhere from -15 to -11, corresponding to 
a $SU(5)$ condensate scale range of 
$1\times 10^{14}$ GeV to $2\times 10^{15}$ GeV.  
Similarly, $\beta_3$ could be anywhere from -9 to -5, corresponding to 
a $SU(3)$ 
condensate scale range of $1\times 10^{9}$ GeV to $2\times 10^{13}$ GeV.  
Thus, when generated by hidden sector condensation,
the intermediate mass scale for the MSSM exotics could be anywhere from
$10^{9}$ GeV to $10^{15}$ GeV.

\section{Concluding Comments}

In this letter we have discussed the possibility of achieving optical
unification within a [321] heterotic string. 
Optical unification, recently suggested by J. Giedt \cite{giedt02a}, 
would explain  
the apparent MSSM unification scale near $2.5 \times 10^{16}$ GeV
as a ``virtual image effect'' of an actual unification of couplings
at the string scale, $\sim 5 \times 10^{17}$ GeV. 
An intriguing aspect of optical unification
is that the apparent MSSM unification is not accidental. 
Rather, in this case, 
like the guaranteed appearance of a virtual image between    
a diverging lens and a real object, 
a MSSM unification scale will always appear 
between the intermediate mass scale of the MSSM exotics  
and the string unification scale, 
when coupling strengths are run upward from their measured values at
low energy scales, under the assumption of an intermediate scale desert.   
As movement of the diverging lens or of the real object simply alters 
the position of the virtual image, so too movement of the intermediate mass 
scale or of the string scale simply alters the location of 
the predicted MSSM unification scale. 
Relatedly, detection of an intermediate exotic MSSM mass scale
in combination with the currently extrapolated MSSM unification scale 
would, thus, reveal the string scale. 

We presented herein a model possessing the potential for optical 
unification. This model is of free fermionic construction in the NAHE class. 
Its basis vectors first appeared in \cite{af3,af3a,af3b}.
The differences between this model and those in \cite{af3,af3a,af3b} 
result from a few changes in GSO phases. 
These changes produce a set of MSSM exotic states with properties
strongly suggesting that 
optical unification may be possible within some regions 
of the parameter space of NAHE-based weakly coupled heterotic strings. 
Further research will reveal if this is indeed so \cite{cfmnpw}.

\section{Acknowledgments}
This letter is a product of the 2002 NSF  
High School Summer Science Research 
(HSSSR) and Research Experiences for Undergraduates (REU) 
programs sponsored by the Center for Astrophysics,
Space Physics, and Engineering Research (CASPER) at Baylor University.
Research funding for Viren Desai was provided by the 
HSSSR program; Research funding for 
Heather Hanson, David Robbins, and Scot Shields
was provided by the REU program. 
G.C. thanks Alon Faraggi and Dimitri Nanopoulos 
for numerous helpful discussions regarding NAHE-based models and 
thanks Joel Giedt for helpful discussions regarding optical unification.

\newpage
%======================== REFERENCES =====================================
%========================================================================
%          MACROS FOR REFERENCES
%========================================================================
\def\AEF{A.E. Faraggi}
\def\AP#1#2#3{{\it Ann.\ Phys.}\/ {\bf#1} (#2) #3}
\def\NPB#1#2#3{{\it Nucl.\ Phys.}\/ {\bf B#1} (#2) #3}
\def\NPBPS#1#2#3{{\it Nucl.\ Phys.}\/ {{\bf B} (Proc. Suppl.) {\bf #1}} (#2) 
 #3}
\def\PLB#1#2#3{{\it Phys.\ Lett.}\/ {\bf B#1} (#2) #3}
\def\PRD#1#2#3{{\it Phys.\ Rev.}\/ {\bf D#1} (#2) #3}
\def\PRL#1#2#3{{\it Phys.\ Rev.\ Lett.}\/ {\bf #1} (#2) #3}
\def\PRT#1#2#3{{\it Phys.\ Rep.}\/ {\bf#1} (#2) #3}
\def\PTP#1#2#3{{\it Prog.\ Theo.\ Phys.}\/ {\bf#1} (#2) #3}
\def\MODA#1#2#3{{\it Mod.\ Phys.\ Lett.}\/ {\bf A#1} (#2) #3}
\def\MPLA#1#2#3{{\it Mod.\ Phys.\ Lett.}\/ {\bf A#1} (#2) #3}
\def\IJMP#1#2#3{{\it Int.\ J.\ Mod.\ Phys.}\/ {\bf A#1} (#2) #3}
\def\IJMPA#1#2#3{{\it Int.\ J.\ Mod.\ Phys.}\/ {\bf A#1} (#2) #3}
\def\nuvc#1#2#3{{\it Nuovo Cimento}\/ {\bf #1A} (#2) #3}
\def\RPP#1#2#3{{\it Rept.\ Prog.\ Phys.}\/ {\bf #1} (#2) #3}
\def\etal{{\it et al\/}}
%=========================================================================
%atbib
               
%\bigskip
%\medskip

\def\bibiteml#1#2{ }
\bibliographystyle{unsrt}

\appendix
\def\s{\phantom{-}}

\begin{flushleft}
\begin{table}
{\rm \large\bf Variations on [321] NAHE-Based Models.}
\begin{eqnarray*}
\begin{tabular}{|l|l|l|}
\hline
Model Name & Gauge Group & MSSM Exotic Matter Content\\
\hline
\hline
FCREU1 & $SU(3)_C \times SU(2)_L \times U(1)_Y \times U(1)_{Z'}$ 
              & three $(3,1)_{-  1/3}$, three $(\bar{3},1)_{1/3}$\\
 &$\times U(1)_A \times \prod_{i=1}^{5} U(1)_i$&   one $(3,1)_{1/6}$, one   $(\bar{3},1)_{-1/6}$  \\ 
 &$\times SU(5)_H \times SU(3)_H$& six   $(1,2)_0$        \\
 &$\times U(1)_6\times U(1)_7$& one   $(1,1)_{1/2, Q_A=0}$, one $(1,1)_{-1/2, Q_A= 0}$\\ 
 && three $(1,1)_{-  1/2, Q_A=-  1/2}$, three $(1,1)_{1/2, Q_A=1/2}$\\ 
 && three $(1,1)_{-  1/2, Q_A=1/2}$, three $(1,1)_{1/2, Q_A=-1/2}$\\ 
 && three of $h_{i=1,2,3,4}$, 
    three of $\bar{h}_{i=1,2,3,4}$\\ 
\hline
FCREU2 & $SU(3)_C \times SU(2)_L \times SU(2)_C \times U(1)_{Y'}$
       &   two $(3,1)_{-  1/3}$, two   $(\bar{3},1)_{1/3}$\\
 &$\times U(1)_A \times \prod_{i=1}^{5} U(1)_i$&   one $(3,1)_{1/6}$, 
                                     one $(\bar{3},1)_{-1/6}$  \\ 
 &$\times  SU(3)_H \times SU(2)_{H_1} \times  SU(2)_{H_2}$
           & three $(1,2)_0$, three $(1,2,2_c)_0$              \\
 &$\times \prod_{i=6}^{9} U(1)_i$
  & three $(1,1)_{-  1/2, Q_A= -1/2}$, three $(1,1)_{1/2, Q_A= -1/2}$ \\ 
 && two of ${h}_{i=1,2,3}$, two of $\bar{h}_{i=1,2,3}$         \\ 
\hline
FCREU3 & $SU(3)_C \times SU(2)_L \times U(1)_{Y}\times U(1)_{Z'}$           
  &   one $(3,1)_{-  1/3}$, one   $(\bar{3},1)_{1/3}$\\
 &$\times U(1)_A \times \prod_{i=1}^{5} U(1)_i$ &   one $(3,1)_{1/6}$, one   $(\bar{3},1)_{-1/6}$  \\ 
 &$\times SU(3)_H \times SU(2)_{H_1}\times SU(2)_{H_2}$
 & six   $(1,2)_0$                                   \\
 &$\times \prod_{i=6}^{9} U(1)_i$
 & one   $(1,1)_{-  1/2, Q_A=0}$, one $(1,1)_{1/2, Q_A=0}$ \\ 
 && three $(1,1)_{-  1/2, Q_A=1/2}$, three $(1,1)_{1/2, Q_A=-1/2}$ \\ 
 && three $(1,1)_{-  1/2, Q_A=- 1/2}$,  three $(1,1)_{1/2, Q_A=1/2}$\\ 
 && three of ${h}_{i=1,2,3,4}$, 
    three of $\bar{h}_{i=1,2,3,4}$\\ 
\hline
FCREU4 & $SU(4)_C \times SU(2)_L \times U(1)_{Y'}$ 
  & no $(3,1)$, no $(\bar{3},1)$                     \\
 &$\times U(1)\times \prod_{i=1}^{5} U(1)_i$& three $(1,2)_{0}$ \\
 &$\times SU(5)_H \times SU(3)_H$ 
           & three $(1,2)_{1/2}$, three $(1,2)_{-1/2}$     \\ 
 &$\times U(1)_6\times U(1)_7$
  & one   $(1,1)_{1/2, Q_A=0}$, one  $(1,1)_{-1/2, Q_A=0}$ \\ 
 && three $( 1,1)_{1/2, Q_A=-1/2}$, three  $(1,1)_{-1/2, Q_A=-1/2}$\\ 
 && two of ${h}_{i=1,2,3}$, two of $\bar{h}_{i=1,2,3}$        \\ 
\hline
FCREU5 & $SU(4)_C \times SU(2)_L \times U(1)_{Y'}$ 
  & no $(3,1)$, no $(\bar{3},1)$                     \\
 &$\times U(1)_A \times \prod_{i=1}^{5} U(1)_i$& three $(1,2)_{0}$ \\
 &$\times SU(3)_H \times SU(2)_{H_1}  
            \times SU(2)_{H_2}$& 
    two   $(1,2)_{1/2}$, two $(1,2)_{-1/2}$        \\ 
 &$\times \prod_{i=6}^{9} U(1)_i$
           & one   $(1,1)_{1/2, Q_A=0}$, one $(1,1)_{-1/2, Q_A= 0}$\\ 
 && three $(1,1)_{-  1/2, Q_A=-  1/2}$, three $(1,1)_{1/2, Q_A=- 1/2}$\\ 
 && two of ${h}_{i=1,2,3}$, two of $\bar{h}_{i=1,2,3}$       \\ 
\hline
\end{tabular}
\label{modelsa}
\end{eqnarray*}\\
{Table A1: Gauge Groups and MSSM Exotics Fields of Models. 
The second column gives the gauge group for each model.
The third column entry specifies the MSSM-charged exotic matter content.
An exotic's representation under $SU(3)_C \times SU(2)_L$
is specified by the two numbers in brackets. 
Hypercharge is given by the first subscript and 
anomalous $U(1)$ charge is given for non-Abelian singlets by a second subscript.} 
\end{table}
\end{flushleft}

\begin{flushleft}
\begin{table}
{\rm \large\bf GSO Phases Variations}
\begin{eqnarray*}
\begin{tabular}{|l|l|}
\hline
Model Name & GSO Phase Variations\\
\hline
FCREU1 & 
$C\left({\mb_1\atop \mb_1}\right)=C\left({\gamma\atop \bone}\right)= -C\left({\gamma\atop \alpha, \beta}\right)= 1$
\\ 
\hline
FCREU2 &
$C\left({\mb_1\atop \mb_1}\right)=C\left({\gamma\atop \bone}\right)= 
-C\left({\gamma\atop \alpha, \beta}\right)= -C\left({\beta\atop \bone}\right)= 1$
\\
\hline
FCREU3 &
$C\left({\gamma\atop \bone}\right)= 
-C\left({\gamma\atop \alpha, \beta}\right)= -C\left({\beta\atop \bone}\right)= 1$
\\
\hline
FCREU4 &
$C\left({\gamma\atop \bone}\right)= 
-C\left({\gamma\atop \alpha, \beta}\right)= 1$
\\
\hline
FCREU5 &
$C\left({\gamma\atop \bone}\right)= 
-C\left({\gamma\atop \alpha, \beta}\right)= -C\left({\alpha\atop \bone}\right)= 1$
\\
\hline
\end{tabular}
\label{modelsb}
\end{eqnarray*}\\
{Table A2: New Models and Their GSO Phase Variations. Only the GSO phases differing from those
of \cite{af3} are given.}
\end{table}
\end{flushleft}

\textwidth=7.5in
\oddsidemargin=-18mm
\topmargin=-5mm
\renewcommand{\baselinestretch}{1.3}
\smallskip

\begin{flushleft}
\begin{table}
{\rm \large\bf Model~FCREU1~Fields}
\begin{eqnarray*}
\begin{tabular}{|l||c|cccccccc|c|cc|}
\hline
%Heading Line 1
  $F$      & $(SU(3)_C,$ & $Q_{Y}$ & $Q_{Z'}$  
   & $Q_A$ & $Q_{1}$ & $Q_{2}$ & $Q_{3}$         & $Q_{4}$ 
           & $Q_{5}$ & $(SU(5)_{H},$   & $Q_{6}$ & $Q_{7}$  \\
           & $SU(2)_L)$ & & & & & & & & & $SU(3)_{H})$ & &  \\
\hline
%   F      & 32             &  Y  &  Z  & A & 1  & 2 & 3 & 4  & 5 & 53      & 6 & 7\\
  $Q_{1}$  & $(3,2)$        & 1/6 & 1/6 &1/2&-1/2&1/2& 0 & 1/2& 0 & $(1,1)$ & 0 & 0\\
  $u_{1}$  & $({\bar 3},1)$ &-2/3 & 1/3 &1/2&-1/2&1/2& 0 &-1/2& 0 & $(1,1)$ & 0 & 0\\
  $d_{1}$  & $({\bar 3},1)$ & 1/3 &-2/3 &1/2&-1/2&1/2& 0 &-1/2& 0 & $(1,1)$ & 0 & 0\\
  $L_{1}$  & $(1,2)$        &-1/2 &-1/2 &1/2&-1/2&1/2& 0 & 1/2& 0 & $(1,1)$ & 0 & 0\\
  $e_{1}$  & $(1,1)$        & 1   & 0   &1/2&-1/2&1/2& 0 &-1/2& 0 & $(1,1)$ & 0 & 0\\ 
  $N_{1}$  & $(1,1)$        & 0   & 1   &1/2&-1/2&1/2& 0 &-1/2& 0 & $(1,1)$ & 0 & 0\\
\hline
  $Q_{2}$  & $(3,2)$        & 1/6 & 1/6 &1/2&1/2&1/2&-1/2& 0 & 0 & $(1,1)$ & 0 & 0\\
  $u_{2}$  & $({\bar 3},1)$ &-2/3 & 1/3 &1/2&1/2&1/2& 1/2& 0 & 0 & $(1,1)$ & 0 & 0\\
  $d_{2}$  & $({\bar 3},1)$ & 1/3 &-2/3 &1/2&1/2&1/2& 1/2& 0 & 0 & $(1,1)$ & 0 & 0\\
  $L_{2}$  & $(1,2)$        &-1/2 &-1/2 &1/2&1/2&1/2&-1/2& 0 & 0 & $(1,1)$ & 0 & 0\\
  $e_{2}$  & $(1,1)$        & 1   & 0   &1/2&1/2&1/2& 1/2& 0 & 0 & $(1,1)$ & 0 & 0\\ 
  $N_{2}$  & $(1,1)$        & 0   & 1   &1/2&1/2&1/2& 1/2& 0 & 0 & $(1,1)$ & 0 & 0\\
\hline
  $Q_{3}$  & $(3,2)$        & 1/6 & 1/6 &1/2& 0 &-1 & 0 & 0 &-1/2& $(1,1)$ & 0 & 0\\
  $u_{3}$  & $({\bar 3},1)$ &-2/3 & 1/3 &1/2& 0 &-1 & 0 & 0 & 1/2& $(1,1)$ & 0 & 0\\
  $d_{3}$  & $({\bar 3},1)$ & 1/3 &-2/3 &1/2& 0 &-1 & 0 & 0 & 1/2& $(1,1)$ & 0 & 0\\
  $L_{3}$  & $(1,2)$        &-1/2 &-1/2 &1/2& 0 &-1 & 0 & 0 &-1/2& $(1,1)$ & 0 & 0\\
  $e_{3}$  & $(1,1)$        & 1   & 0   &1/2& 0 &-1 & 0 & 0 & 1/2& $(1,1)$ & 0 & 0\\ 
  $N_{3}$  & $(1,1)$        & 0   & 1   &1/2& 0 &-1 & 0 & 0 & 1/2& $(1,1)$ & 0 & 0\\
\hline
% higgs-like doublets
      $h_1$& $(1,2)$        &-1/2 & 1/2 & 1  &-1  & 1  & 0 & 0 & 0 & $(1,1)$ & 0 & 0\\
      $h_2$& $(1,2)$        &-1/2 & 1/2 & 1  & 1  & 1  & 0 & 0 & 0 & $(1,1)$ & 0 & 0\\
      $h_3$& $(1,2)$        &-1/2 & 1/2 & 1  & 0  &-2  & 0 & 0 & 0 & $(1,1)$ & 0 & 0\\
      $h_4$& $(1,2)$        &-1/2 & 0   &-1/4&-1/2&1/2 & 0 & 0 & 0 & $(1,1)$ & 2 & 0\\
$\bar{h}_1$& $(1,2)$        & 1/2 &-1/2 &-1  & 1  &-1  & 0 & 0 & 0 & $(1,1)$ & 0 & 0\\
$\bar{h}_2$& $(1,2)$        & 1/2 &-1/2 &-1  &-1  &-1  & 0 & 0 & 0 & $(1,1)$ & 0 & 0\\
$\bar{h}_3$& $(1,2)$        & 1/2 &-1/2 &-1  & 0  & 2  & 0 & 0 & 0 & $(1,1)$ & 0 & 0\\
$\bar{h}_4$& $(1,2)$        & 1/2 & 0   &1/4 &1/2 &-1/2& 0 & 0 & 0 & $(1,1)$ &-2 & 0\\
\hline
% exotic triplets
      $D_1$& $(3,1)$        &-1/3 &-1/3 & 1  & 0  & 1  & 0  &  0 & 0  & $(1,1)$ & 0 & 0 \\
      $D_2$& $(3,1)$        &-1/3 &-1/3 &-1  & 0  &-1  & 0  &  0 & 0  & $(1,1)$ & 0 & 0 \\
      $D_3$& $(3,1)$        &-1/3 & 1/6 & 1/4&-1/2&-1/2& 0  &  0 & 0  & $(1,1)$ &-2 & 0 \\
      $D_4$& $(3,1)$        & 1/6 & 1/6 & 0  & 0  & 0  &1/2 & 1/2& 1/2& $(1,1)$ &1/2&-15/2\\
$\bar{D}_1$& $({\bar 3},1)$ & 1/3 & 1/3 &-1  & 0  &-1  & 0  &  0 & 0  & $(1,1)$ & 0 & 0 \\
$\bar{D}_2$& $({\bar 3},1)$ & 1/3 & 1/3 & 1  & 0  & 1  & 0  &  0 & 0  & $(1,1)$ & 0 & 0 \\
$\bar{D}_3$& $({\bar 3},1)$ & 1/3 & 1/6 &-1/4&1/2 &1/2 & 0  &  0 & 0  & $(1,1)$ & 2 & 0 \\
$\bar{D}_4$& $({\bar 3},1)$ &-1/6 &-1/6 & 0  & 0  & 0  &-1/2&-1/2&-1/2& $(1,1)$ &-1/2& 15/2\\
\hline
% exotic doublets
      $X_1$& $(1,2)$        & 0   & 0   & 1/2&-1/2& 1/2&1/2& 0  &1/2& $(1,1)$ &-1/2& 15/2\\
      $X_2$& $(1,2)$        & 0   & 0   & 1/2& 1/2& 1/2& 0 &-1/2&1/2& $(1,1)$ &-1/2& 15/2\\
      $X_3$& $(1,2)$        & 0   & 0   & 1/2& 0  &-1  &1/2&-1/2& 0 & $(1,1)$ &-1/2& 15/2\\
$\bar{X}_1$& $(1,2)$        & 0   & 0   &-1/2& 1/2&-1/2&1/2& 0  &1/2& $(1,1)$ & 1/2&-15/2\\
$\bar{X}_2$& $(1,2)$        & 0   & 0   &-1/2&-1/2&-1/2& 0 &-1/2&1/2& $(1,1)$ & 1/2&-15/2\\
$\bar{X}_3$& $(1,2)$        & 0   & 0   &-1/2& 0  & 1  &1/2&-1/2& 0 & $(1,1)$ & 1/2&-15/2\\
\hline
\end{tabular}
\nolabel
\end{eqnarray*}
Table A3: Model FCREU1 Fields and Their Charges.
\end{table}
\end{flushleft}
   
\begin{flushleft}
\begin{table}
%{\rm \large\bf Model~FCREU1~Fields}
\begin{eqnarray*}
\begin{tabular}{|l||c|cccccccc|c|cc|}
\hline
%Heading Line 1
  $F$      & $(SU(3)_C,$ & $Q_{Y}$ & $Q_{Z'}$  
   & $Q_A$ & $Q_{1}$ & $Q_{2}$ & $Q_{3}$         & $Q_{4}$ 
           & $Q_{5}$ & $(SU(5)_{H},$   & $Q_{6}$ & $Q_{7}$  \\
           & $SU(2)_L)$ & & & & & & & & & $SU(3)_{H})$ & &  \\
\hline
%   F      & 32             &  Y  &  Z  & A & 1  & 2 & 3 & 4  & 5 & 53      & 6 & 7\\
% exotic singlets
      $A_1$& $(1,1)$        & 1/2 & 1/2 & 0  & 0  & 0  & 1/2& 1/2&-1/2& $(1,1)$ &-1/2& 15/2\\
      $A_2$& $(1,1)$        &-1/2 & 1/2 &-1/2&-1/2&-1/2& 0  & 1/2&-1/2& $(1,1)$ &-1/2& 15/2\\
      $A_3$& $(1,1)$        &-1/2 & 1/2 &-1/2& 0  & 1  &-1/2& 1/2& 0  & $(1,1)$ &-1/2& 15/2\\
      $A_4$& $(1,1)$        &-1/2 & 1/2 &-1/2& 1/2&-1/2&-1/2& 0  &-1/2& $(1,1)$ &-1/2& 15/2\\
      $A_5$& $(1,1)$        & 1/2 &-1/2 &-1/2&-1/2&-1/2& 0  & 1/2&-1/2& $(1,1)$ &-1/2& 15/2\\
      $A_6$& $(1,1)$        & 1/2 &-1/2 &-1/2& 0  & 1  &-1/2& 1/2& 0  & $(1,1)$ &-1/2& 15/2\\
      $A_7$& $(1,1)$        & 1/2 &-1/2 &-1/2& 1/2&-1/2&-1/2& 0  &-1/2& $(1,1)$ &-1/2& 15/2\\
$\bar{A}_1$& $(1,1)$        &-1/2 &-1/2 & 0  & 0  & 0  &-1/2&-1/2& 1/2& $(1,1)$ & 1/2&-15/2\\
$\bar{A}_2$& $(1,1)$        & 1/2 &-1/2 & 1/2& 1/2& 1/2& 0  & 1/2&-1/2& $(1,1)$ & 1/2&-15/2\\
$\bar{A}_3$& $(1,1)$        & 1/2 &-1/2 & 1/2& 0  &-1  &-1/2& 1/2& 0  & $(1,1)$ & 1/2&-15/2\\
$\bar{A}_4$& $(1,1)$        & 1/2 &-1/2 & 1/2&-1/2& 1/2&-1/2& 0  &-1/2& $(1,1)$ & 1/2&-15/2\\
$\bar{A}_5$& $(1,1)$        &-1/2 & 1/2 & 1/2& 1/2& 1/2& 0  & 1/2&-1/2& $(1,1)$ & 1/2&-15/2\\
$\bar{A}_6$& $(1,1)$        &-1/2 & 1/2 & 1/2& 0  &-1  &-1/2& 1/2& 0  & $(1,1)$ & 1/2&-15/2\\
$\bar{A}_7$& $(1,1)$        &-1/2 & 1/2 & 1/2&-1/2& 1/2&-1/2& 0  &-1/2& $(1,1)$ & 1/2&-15/2\\
\hline
% VEV singlets
  $\Phi_1$ & $(1,1)$        & 0   & 0   & 0 & 0 & 0 & 0 & 0 & 0 & $(1,1)$ & 0 & 0\\
  $\Phi_2$ & $(1,1)$        & 0   & 0   & 0 & 0 & 0 & 0 & 0 & 0 & $(1,1)$ & 0 & 0\\
  $\Phi_3$ & $(1,1)$        & 0   & 0   & 0 & 0 & 0 & 0 & 0 & 0 & $(1,1)$ & 0 & 0\\
$\Phi_{12}$& $(1,1)$        & 0   & 0   & 0 &-2 & 0 & 0 & 0 & 0 & $(1,1)$ & 0 & 0\\
$\Phi_{23}$& $(1,1)$        & 0   & 0   & 0 & 1 &-3 & 0 & 0 & 0 & $(1,1)$ & 0 & 0\\
$\Phi_{31}$& $(1,1)$        & 0   & 0   & 0 &-1 &-3 & 0 & 0 & 0 & $(1,1)$ & 0 & 0\\
${\bar\Phi}_{3}$& $(1,1)$   & 0   & 0   & 0 & 2 & 0 & 0 & 0 & 0 & $(1,1)$ & 0 & 0\\
${\bar\Phi}_{23}$&$(1,1)$   & 0   & 0   & 0 &-1 & 3 & 0 & 0 & 0 & $(1,1)$ & 0 & 0\\
${\bar\Phi}_{31}$&$(1,1)$   & 0   & 0   & 0 & 1 & 3 & 0 & 0 & 0 & $(1,1)$ & 0 & 0\\
      $S_1$& $(1,1)$        & 0   & 0   & 0 &-1 & 0 &-1 & 0 & 0 & $(1,1)$ & 0 & 0\\
      $S_2$& $(1,1)$        & 0   & 0   & 0 &-1 & 0 & 1 & 0 & 0 & $(1,1)$ & 0 & 0\\
      $S_3$& $(1,1)$        & 0   & 0   & 0 &-1 & 0 & 0 &-1 & 0 & $(1,1)$ & 0 & 0\\
      $S_4$& $(1,1)$        & 0   & 0   & 0 &-1 & 0 & 0 & 1 & 0 & $(1,1)$ & 0 & 0\\
      $S_5$& $(1,1)$        & 0   & 0   & 0 &-1 & 0 & 0 & 0 &-1 & $(1,1)$ & 0 & 0\\
      $S_6$& $(1,1)$        & 0   & 0   & 0 &-1 & 0 & 0 & 0 & 1 & $(1,1)$ & 0 & 0\\
      $S_7$& $(1,1)$        & 0   & 1/2 &3/4&-1/2&-3/2&0& 0 & 0 & $(1,1)$ & 2 & 0\\
      $S_8$& $(1,1)$        & 0   & 1/2 &3/4&1/2&3/2& 0 & 0 & 0 & $(1,1)$ & 2 & 0\\
      $S_9$& $(1,1)$        & 0   & 1/2 &-5/4&1/2&-1/2&0& 0 & 0 & $(1,1)$ & 2 & 0\\
$\bar{S}_1$& $(1,1)$        & 0   & 0   & 0 & 1 & 0 & 1 & 0 & 0 & $(1,1)$ & 0 & 0\\
$\bar{S}_2$& $(1,1)$        & 0   & 0   & 0 & 1 & 0 &-1 & 0 & 0 & $(1,1)$ & 0 & 0\\
$\bar{S}_3$& $(1,1)$        & 0   & 0   & 0 & 1 & 0 & 0 & 1 & 0 & $(1,1)$ & 0 & 0\\
$\bar{S}_4$& $(1,1)$        & 0   & 0   & 0 & 1 & 0 & 0 &-1 & 0 & $(1,1)$ & 0 & 0\\
$\bar{S}_5$& $(1,1)$        & 0   & 0   & 0 & 1 & 0 & 0 & 0 & 1 & $(1,1)$ & 0 & 0\\
$\bar{S}_6$& $(1,1)$        & 0   & 0   & 0 & 1 & 0 & 0 & 0 &-1 & $(1,1)$ & 0 & 0\\
$\bar{S}_7$& $(1,1)$        & 0   &-1/2 &-3/4&1/2&3/2&0 & 0 & 0 & $(1,1)$ &-2 & 0\\
$\bar{S}_8$& $(1,1)$        & 0   &-1/2 &-3/4 &-1/2&-3/2& 0 & 0 & 0 & $(1,1)$ &-2 & 0\\
$\bar{S}_9$& $(1,1)$        & 0   &-1/2 & 5/4 &-1/2& 1/2& 0 & 0 & 0 & $(1,1)$ &-2 & 0\\
\hline
\end{tabular}
\nolabel
\end{eqnarray*}
Table A3 continued: Model FCREU1 Fields and Their Charges.
\end{table}
\end{flushleft}
   
\begin{flushleft}
\begin{table}
%{\rm \large\bf Model~FCREU1~Fields}
\begin{eqnarray*}
\begin{tabular}{|l||c|cccccccc|c|cc|}
\hline
%Heading Line 1
  $F$      & $(SU(3)_C,$ & $Q_{Y}$ & $Q_{Z'}$  
   & $Q_A$ & $Q_{1}$ & $Q_{2}$ & $Q_{3}$         & $Q_{4}$ 
           & $Q_{5}$ & $(SU(5)_{H},$   & $Q_{6}$ & $Q_{7}$  \\
           & $SU(2)_L)$ & & & & & & & & & $SU(3)_{H})$ & &  \\
\hline
%   F      & 32             &  Y  &  Z  & A & 1  & 2 & 3 & 4  & 5 & 53      & 6 & 7\\
% hidden sector 5's
      $F_1$& $(1,1)$        & 0   &-1/2 &-1/4&-1/2& 1/2& 0  & 0  & 0  &$(5,1)$ &-1  &-3\\
      $F_2$& $(1,1)$        & 0   & 0   & 1  &  0 &  1 & 0  & 0  &1/2 &$(5,1)$ & 1  &-3\\
      $F_3$& $(1,1)$        & 0   & 0   & 1  &-1/2&-1/2& 1/2& 0  & 0  &$(5,1)$ & 1  &-3\\
      $F_4$& $(1,1)$        & 0   & 0   & 1  & 1/2&-1/2& 0  &-1/2& 0  &$(5,1)$ & 1  &-3\\

$\bar{F}_1$& $(1,1)$        & 0   & 1/2 & 1/4& 1/2&-1/2& 0  & 0  & 0  &$({\bar 5},1)$ & 1  & 3\\
$\bar{F}_2$& $(1,1)$        & 0   & 0   & 1  &  0 &  1 & 0  & 0  &-1/2&$({\bar 5},1)$ &-1  & 3\\
$\bar{F}_3$& $(1,1)$        & 0   & 0   & 1  &-1/2&-1/2&-1/2& 0  & 0  &$({\bar 5},1)$ &-1  & 3\\
$\bar{F}_4$& $(1,1)$        & 0   & 0   & 1  & 1/2&-1/2& 0  & 1/2& 0  &$({\bar 5},1)$ &-1  & 3\\
\hline
% hidden sector 3's
      $K_1$& $(1,1)$        & 0   & 1/2 & 1/4&-1/2&-1/2& 0  &  0 & 0  &$(1,3)$ & 1  &-5\\
      $K_2$& $(1,1)$        & 0   & 0   & 1  & 1/2&-1/2& 0  &-1/2& 0  &$(1,3)$ &-1  &-5\\
      $K_3$& $(1,1)$        & 0   & 0   & 1  &-1/2&-1/2& 1/2&  0 & 0  &$(1,3)$ &-1  &-5\\
      $K_4$& $(1,1)$        & 0   & 0   & 1  &  0 & 1  & 0  &  0 & 1/2&$(1,3)$ &-1  &-5\\
$\bar{K}_1$& $(1,1)$        & 0   &-1/2 &-1/4& 1/2&-1/2& 0  &  0 & 0  &$(1,{\bar 3})$ &-1  & 5\\
$\bar{K}_2$& $(1,1)$        & 0   & 0   & 1  & 1/2&-1/2& 0  & 1/2& 0  &$(1,{\bar 3})$ & 1  & 5\\
$\bar{K}_3$& $(1,1)$        & 0   & 0   & 1  &-1/2&-1/2&-1/2&  0 & 0  &$(1,{\bar 3})$ & 1  & 5\\
$\bar{K}_4$& $(1,1)$        & 0   & 0   & 1  &  0 & 1  & 0  &  0 &-1/2&$(1,{\bar 3})$ & 1  & 5\\
\hline
\end{tabular}
\nolabel
\end{eqnarray*}
Table A3 continued: Model FCREU1 Fields and Their Charges.
\end{table}
\end{flushleft}

%==============================================================================
\end{document}